\documentclass[lettersize,journal]{IEEEtran}

\usepackage{color}
\usepackage[hyphens]{url}
\usepackage{multirow}
\usepackage{booktabs,makecell}
\usepackage{adjustbox}
\usepackage{enumitem}
\usepackage{cite}
\usepackage{amsmath,amssymb,amsfonts}
\usepackage{algorithmic}
\usepackage{array}
\usepackage[caption=false,font=normalsize,labelfont=sf,textfont=sf]{subfig}
\usepackage{textcomp}
\usepackage{stfloats}
\usepackage{url}
\usepackage{verbatim}
\usepackage{graphicx}

\usepackage{longtable}

\usepackage[T1]{fontenc} 
\usepackage[utf8]{inputenc} 

\usepackage{lineno}

\def\BibTeX{{\rm B\kern-.05em{\sc i\kern-.025em b}\kern-.08em
    T\kern-.1667em\lower.7ex\hbox{E}\kern-.125emX}}
\usepackage{balance}
\begin{document}
\title{Are We There Yet? A Brief Survey of Music Emotion Prediction Datasets, Models and Outstanding Challenges}


\author{Jaeyong Kang, Dorien Herremans \IEEEmembership{Senior Member, IEEE}
\thanks{J. Kang and D. Herremans are with the Information Systems Technology and Design Pillar at the Singapore University of Technology and Design (SUTD), Singapore 487372. E-mail: \{jaeyong\_kang, dorien\_herremans\}@sutd.edu.sg.}
\thanks{J. Kang is the corresponding author. }
}
      
\markboth{Journal of \LaTeX\ Class Files, October~2024}%
{How to Use the IEEEtran \LaTeX \ Templates}

\maketitle

\begin{abstract}
Deep learning models for music have advanced drastically in recent years, but how good are machine learning models at capturing emotion, and what challenges are researchers facing? In this paper, we provide a comprehensive overview of the available music-emotion datasets and discuss evaluation standards as well as competitions in the field. We also offer a brief overview of various types of music emotion prediction models that have been built over the years, providing insights into the diverse approaches within the field. Through this examination, we highlight the challenges that persist in accurately capturing emotion in music, including issues related to dataset quality, annotation consistency, and model generalization. Additionally, we explore the impact of different modalities, such as audio, MIDI, and physiological signals, on the effectiveness of emotion prediction models. Through this examination, we identify persistent challenges in music emotion recognition (MER), including issues related to dataset quality, the ambiguity in emotion labels, and the difficulties of cross-dataset generalization. We argue that future advancements in MER require standardized benchmarks, larger and more diverse datasets, and improved model interpretability. Recognizing the dynamic nature of this field, we have complemented our findings with an accompanying GitHub repository\footnote{ \url{https://github.com/AMAAI-Lab/awesome-MER/}}. This repository contains a comprehensive list of music emotion datasets and recent predictive models.
\end{abstract}


\section{Introduction}
Music has long been revered for its profound ability to evoke and convey emotions, transcending cultural, linguistic, and geographical barriers. Researchers from various fields have been captivated by the intricate interplay between music and human emotions for decades. Classic works such as Meyer's paper \cite{leonard1956emotion} and pioneering studies conducted by scholars such as Seashore \cite{seashore1937psychology} and Hevner \cite{hevner1935affective} in the early 20th century laid the groundwork for understanding the emotional impact of music. However, understanding and quantifying this poses a significant challenge due to its multifaceted nature \cite{juslin2008emotional}. In this paper, we provide an overview of the current state-of-the-art along with challenges and directions for future work. 

With the advent of technology and data-driven methodologies, new avenues have opened up for exploring the complex relationship between music and emotions, such as deep learning models. Despite these advancements, accurate predictions of emotions from music remain elusive. This is due to a number of reasons, including the subjective, personal nature of emotional perception, as well as bias and limitations in the current datasets, and challenges in benchmarking. We discuss these and other challenges at length in Section~\ref{sec:challenges}. 

Despite these challenges, the potential applications of Music Emotion Recognition (MER) systems are vast and varied. In the healthcare domain, we may find personalized music recommendation systems that guide a listener to different emotional states \cite{agres2021music}, or we may even see MER systems used to build large-scale datasets on which we can train generative music AI systems that can be controlled to generate music with specific emotions \cite{makris2021generating, dash2024ai, herremans2017morpheus}. In the same line of thought, \cite{agres2023affectmachine} developed a Brain-Computer Interface system that can provide musical feedback about the listener's current emotional state and subsequently influence this state. Additionally, MER systems may help facilitate emotional analysis during music composition, interactive experiences in media and entertainment, as well as inform market research \cite{yang2012machine}. The implications of understanding music and emotions extend across diverse domains. In general, following the idea of positive psychology \cite{seligman2000positive}, understanding music emotions can be used to enhance the user experience when designing various systems. 

In this paper, we do not aim to provide a comprehensive overview of MER models, instead, we focus on discussing datasets, evaluation approaches, and identifying key challenges as well as future directions. We only briefly touch upon some of the more recent MER models. For a more comprehensive overview of MER machine learning models, the reader is referred to \cite{han2022survey, kim2010music, yang2018review, yang2012machine, barthet2013music}. 

Earlier surveys on Music Emotion Recognition (MER) include the review by Panda et al. (2020) \cite{panda2020audio}, which focuses on the taxonomy of audio features and their relationship to emotion perception. More recently, a broad overview of deep learning techniques for MER was provided in \cite{jiang2024music}, with extended discussion on emotion representations. However, these surveys primarily focus on either specific feature engineering strategies or model architectures, and they often lack comprehensive coverage of datasets, evaluation protocols, and the challenges of integrating multimodal sources or predicting induced emotion. In contrast, our goal is to provide a broader and more integrative perspective on the MER landscape. We examine not only the evolution of model architectures, but also the diversity of datasets, annotation strategies, emotion models, and evaluation practices. This comprehensive perspective allows us to highlight emerging trends, identify gaps in current methodologies, and suggest promising directions for future research.

In the next section, we provide an extensive overview of the available emotion-annotated music datasets. This is followed by a discussion of the evaluation practices in the field (Section~\ref{sec:eval}). After that, we describe a selected list of recent models and approaches in Section~\ref{sec:5}. Finally, Section~\ref{sec:challenges} dives into the remaining challenges and future direction for the field of music emotion recognition (MER), followed by a general conclusion.

\begin{table*}[ht!]
\small
\centering
\label{tab:datasets}
\caption{Overview of Music Emotion Datasets (Sorted by Year). Note: S = Static, D = Dynamic, B = Both; P = Perceived, I = Induced.}
\resizebox{\textwidth}{!}{ 
\begin{tabular}{@{}lllllllll@{}}
\toprule
Dataset & Year & \# of instances & Length & Type & Categorical & Dimensional & S/D/B & P/I \\ \midrule
MoodsMIREX \cite{hu2007exploring} & 2007 & 269 &30s & MP3 &  5 labels & - & S & P \\
CAL500 \cite{turnbull2007towards} & 2007 & 500 & full & MP3 & 174 labels & - & S & P \\
Yang-Dim \cite{yang2008regression} & 2008 & 195 & 25s & WAV & - & Russell & S & P \\
MoodSwings \cite{kim2008moodswings} & 2008 & 240 & 15s & MP3 & - & Russell & D & P \\
NTWICM \cite{schuller2010determination} & 2010 & 2,648 & full & MP3 & - & Russell & S & P \\
Soundtrack \cite{eerola2011comparison} & 2011 & 470 & 15s-1m & MP3 & 6 labels & 3 dimensions & S & P \\
MoodSwings Turk \cite{speck2011comparative} & 2011 & 240 & 15s & MP3 & - & Russell & D & P \\
Last.fm subset of MSD \cite{bertin2011million} & 2011 & 505,216 & full & Metadata only & listener tags & - & S & P \\
DEAP \cite{koelstra2011deap} & 2012 & 120 & 60s & YouTube id & - & Russell & S & I \\
Panda et al.'s dataset \cite{panda2013multi} & 2013 & 903 & 30s & MP3, MIDI & 21 labels & - & S & P \\
Solymani et al.'s dataset \cite{soleymani20131000} & 2013 & 1,000 & 45s & MP3 & - & Russell & B & P \\
CAL500exp \cite{wang2014towards} & 2014 & 3,223 & 3s-16s & MP3 & 67 labels & - & S & P \\
AMG1608 \cite{chen2015amg1608} & 2015 & 1,608 & 30s & WAV & - & Russell & S & P \\
Emotify \cite{zentner2008emotions} & 2016 & 400 & 60s & MP3 & GEMS & - & S & I \\
Moodo \cite{pesek2017moodo} & 2016 & 200 & 15s & WAV & - & Russell & S & P \\
Malheiro et al.'s dataset \cite{malheiro2016bi} & 2016 & 200 & 30s & Audio, Lyrics & Quadrants & - & S & P \\
CH818 \cite{hu2017mood} & 2017 & 818 & 30s & MP3 & - & Russell & S & P \\
MoodyLyrics \cite{ccano2017moodylyrics} & 2017 & 2,595 & full & Lyrics & 4 labels & - & S & P \\
4Q-emotion \cite{panda2018musical} & 2018 & 900 & 30s & MP3 & Quadrants & - & S & P \\
DEAM \cite{aljanaki2017developing} & 2018 & 2,058 & 45s & MP3 & - & Russell & B & P \\
PMEmo \cite{zhang2018pmemo} & 2018 & 794 & full & MP3 & - & Russell & B & I \\
RAVDESS \cite{livingstone2018ryerson} & 2018 & 1,012 & full & MP3, MP4 & 5 labels & - & S & P \\
DMDD \cite{delbouys2018music} & 2018 & 18,644 & full & Audio, Lyrics & - & Russell & S & P \\
MTG-Jamendo \cite{bogdanov2019mtg} & 2019 & 18,486 & full & MP3 & 56 labels & - & S & P \\
VGMIDI \cite{ferreira2021learning} & 2019 & 200 & full & MIDI & - & Russell & D & P \\
Turkish Music Emotion \cite{er2019music} & 2019 & 400 & 30s & MP3 & 4 labels & - & S & P \\
EMOPIA \cite{hung2021emopia} & 2021 & 1,087 & 30s-40s & Audio, MIDI & Quadrants & - & S & P \\
MER500 \cite{velankar2020} & 2020 & 494 & 10s & WAV & 5 labels & - & S & P \\
Music4all \cite{santana2020music4all} & 2020 & 109,269 & 30s & WAV & - & 3 dimensions & S & P \\
CCMED-WCMED \cite{fan2020comparative} & 2020 & 800 & 8-20s & WAV & - & Russell & S & P \\
MuSe \cite{akiki2021muse} & 2021 & 90,001 & full & Audio & - & Russell (V-A-D) & S & P \\
HKU956 \cite{hu2022detecting} & 2022 & 956 & full & MP3 & - & Russell & S & I \\
MERP \cite{koh2022merp} & 2022 & 54 & full & WAV & - & Russell & B & P \\
MuVi \cite{chua2022predicting} & 2022 & 81 & full & YouTube id & GEMS & Russell & B & P \\
YM2413-MDB \cite{choi2022ym2413} & 2022 & 699 & full & WAV, MIDI & 19 labels & - & S & P \\
MusAV \cite{bogdanov2022musav} & 2022 & 2,092 & full & WAV & - & Russell & S & P \\
EmoMV \cite{thao2023emomv} & 2023 & 5,986 & 30s & WAV & 6 labels & - & S & P \\
Indonesian Song \cite{sams2023multimodal} & 2023 & 476 & full & WAV & 3 labels & - & S & P \\
TROMPA-MER \cite{gomez2023trompa} & 2023 & 1,161 & 30s & WAV & 11 labels & - & S & P \\
Music-Mouv \cite{doumbia2023characterizing} & 2023 & 188 & full & Spotify id & GEMS & - & S & I \\
ENSA \cite{ospitia2023ensa} & 2023 & 60 & full & MP3 & - & Russell & D & P \\
EMMA \cite{strauss2024emotion} & 2024 & 364 & 30s-60s & WAV & GEMS & - & S & I \\
SiTunes \cite{grigorev2024situnes} & 2024 & 300 & full & WAV & - & Russell & S & I \\
MERGE \cite{louro2024merge} & 2024 & 3,554 & full & Audio, Lyrics & Quadrants & - & S & P \\
Popular Hooks \cite{wu2024popular} & 2024 & 38,694 & hooks & Video, Audio, Lyrics & Quadrants & - & S & P \\
Affolter and Rohrmeier's dataset \cite{affolter2024utilizing} & 2024 & 5,892 & full & Spotify id & 8 labels & - & S & P \\
XMIDI \cite{tian2025xmusic} & 2025 & 108,023 & full & MIDI & 11 labels & - & S & P \\
\bottomrule
\end{tabular}
}
\end{table*}

\section{Datasets} 
\label{sec:datasets}
Table \ref{tab:datasets} presents an extensive overview of emotion-annotated music datasets. These datasets vary in size, annotation granularity, and focus on either perceived or induced emotions. We have attempted to provide an exhaustive overview of emotion datasets. This was achieved by using Google Scholar with search terms such as `music emotion dataset', `affective music dataset', as well as following references within these articles. Before delving deeper into the datasets, we first describe key dimensions that help structure our comparison. These include: (1) the emotion representation model (e.g., categorical vs. dimensional), (2) the annotation method (static or dynamic), (3) the type of emotion captured (perceived vs. induced), (4) the data modalities involved (e.g., audio, video, lyrics), and (5) the dataset’s size and diversity in genre or listener demographics. Below, we elaborate on each of these aspects and illustrate them with representative examples.


\textbf{Emotion representations}
One of the earliest emotion representation models in music is Hevner’s affective ring \cite{hevner1936experimental}, developed in 1936. Based on extensive experimental studies, Hevner’s model categorizes music emotions into eight fundamental categories: dignified, sad, dreamy, serene, graceful, happy, exciting, and vigorous. In current-day MER research, we see that Russell’s Circumplex Model of Affect \cite{russell1980circumplex} is widely used. This model characterizes emotions along two dimensions: valence and arousal. Valence represents the degree of positive or negative emotion, while arousal reflects the intensity of emotion, ranging from passive to activated states. Russell’s original model contained a third dimension: dominance \cite{russell1980circumplex}. This dimension is typically omitted as it can be hard to annotate, although some researchers have argued to re-include it \cite{bakker2014pleasure}. It is worth noting, however, that the dominance dimension was more central to Russell’s earlier PAD (Pleasure-Arousal-Dominance) framework, designed for environmental and contextual emotion analysis. Its omission from the Circumplex Model was a simplification intended to facilitate self-reported emotion studies.

Russell’s model is a \textit{dimensional} model, as it consists of continuous values along multiple dimensions (valence/arousal). Hevner’s model, on the other hand, is \textit{categorical} as it consists of discrete emotion labels. Thayer’s two-dimensional model \cite{thayer1990biopsychology} focuses on energetic arousal and tense arousal as the primary dimensions of emotion. Thayer suggests that valence can be inferred from the combination of energetic and tense arousal levels. Other categorical models include the Geneva Emotional Music Scales (GEMS) \cite{zentner2008emotions}. This model was specifically designed for music-induced emotions and consists of 45 emotion tags grouped into nine categories, including amazement, solemnity, tenderness, nostalgia, calmness, power, joyful activation, tension, and sadness.

Recently, alternative emotion frameworks grounded in psychological theory have also been explored. For instance, Affolter et al. \cite{affolter2024utilizing} introduced a dataset that maps listener-generated tags and playlist names to emotion labels derived from Plutchik’s psychoevolutionary model of emotions. Their method employs natural language processing to associate each track with an 8-dimensional vector corresponding to Plutchik’s basic emotions, offering a novel bridge between tag-based annotations and theoretical models of affect.

The datasets listed in Table~\ref{tab:datasets} use a variety of emotion representation models, with Russell’s model being the most popular dimensional representation. Some datasets do not use a specific emotion model, for instance, the MTG-Jamendo dataset \cite{bogdanov2019mtg} consists of freely assigned tags by the listeners, resulting in a diverse and comprehensive set of 56 tags, spanning from melancholic' to upbeat’. A subset of this dataset is used for the Emotion and Theme Recognition in Music' Task of MediaEval \cite{bogdanov2019mediaeval}, which serves as a benchmark for evaluating MER systems. In Table~\ref{tab:datasets}, the number of tags used to represent the emotions are listed in the column ``Categorical''.
 
\textbf{Static versus dynamic} 
Regardless of which emotion representation model is being used, we notice two fundamentally different approaches: static versus dynamic annotations. In a static setting, the listeners indicate the emotion for the entire song or fragment. In a dynamic annotation setting, the listener continuously indicates the emotion throughout the song or fragment. For instance, the MTG-Jamendo dataset \cite{bogdanov2019mtg} offers categorical tags for each full-length song. The annotation is static, but multiple tags are allowed per song. The MoodSwings dataset \cite{kim2008moodswings}, on the other hand, offers dynamic annotations of valence/arousal for every second of the musical fragments. Finally, some datasets offer both, for instance, MERP \cite{koh2022merp} provides both static GEM labels for the entire song, as well as dynamic valence/arousal ratings for every 1s of a song. 

\textbf{Induced versus perceived}
Emotion labels in the datasets can represent either perceived or induced emotions. Perceived emotions are emotions that listeners consciously recognize within the music itself. Induced emotions, on the other hand, are emotions that listeners experience as a result of listening to the music, which involve an actual emotional experience provoked by the stimulus. This emotional experience can be influenced by context, memories, and personal experiences, and may differ from the perceived emotion \cite{juslin2004expression}.

Most of the datasets in Table~\ref{tab:datasets} use perceived emotion labels, which are easier to annotate. For induced emotion labels, researchers have used psycho-physiological measurements ranging from electromyogram (EMG), volume pulse (BVP), electrocardiograms (ECG), skin conductance, respiration rate, heart rate, to electroencephalograms (EEG). There have been many studies in psychology that use such biosensors to explore how  music can influence our emotions \cite{trochidis2013psychophysiological, salimpoor2011anatomically, jaimovich2013emotion, mcfarland1985relationship, kim2008emotion}. Since these studies include medical data, the datasets are not often public. However, there are a few datasets with music and its induced emotions. Firstly the DEAP dataset \cite{koelstra2011deap} includes EEG, facial video recordings, as well as peripheral physiological signals that were recorded while they watched music videos. In addition, they collected perceived emotion ratings in terms of arousal, valence, like/dislike, dominance and familiarity. Second, the HKU956 dataset \cite{hu2022detecting} records five kinds of physiological signals (i.e., heart rate, electrodermal activity, blood volume pulse, inter-beat interval, and skin temperature) of participants as they listen to music, along with reported emotions in the arousal and valence dimensions. Third, the Music-Mouv' dataset \cite{doumbia2023characterizing} investigates the impact of emotional context induced by music on gait initiation, focusing on anticipatory postural adjustments crucial for the elderly and individuals with Parkinson's disease. It includes subjective emotional responses, physiological data from wristbands, and biomechanical data from shoe insoles equipped with sensors collected during and after music listening. Finally, SiTunes \cite{grigorev2024situnes} includes physiological signals (i.e., heart rate, activity intensity, activity step, and activity type) measured both before and after music listening, alongside environmental data (i.e., time of day, weather information, and location) recorded during users' daily lives. 
The MER models that can predict \textit{induced} emotions have various medical applications, such as curating playlists to guide patients to different emotional states \cite{agres2021music}. An interesting study by \cite{song2016perceived} concludes that music tends to induce the emotion that is perceived, enabling researchers to use perceived emotion datasets for developing emotion-inducing models.

\textbf{Modalities}
Music comes in many formats, the most common one being `audio'. Audio files can either be raw waveforms or compressed .mp3 files. However, we should not neglect the MIDI format, a popular format still often used by music producers, composers and performers. Whereas most datasets contain audio files, only a handful focus on MIDI: 1) the VGMIDI dataset \cite{ferreira2021learning}, which contains continuous valence/arousal ratings for MIDI files of piano arrangements for video game soundtracks; 2) the Panda et al.'s dataset \cite{panda2013multi}, which provides a diverse collection of audio clips, lyrics, and aligned MIDI files, contains 28 emotion labels grouped into five emotion clusters derived from a cluster analysis of online tags by \cite{hu2007exploring}; and 3) the EMOPIA dataset \cite{hung2021emopia}, which contains paired piano music audio with MIDI that has emotion annotations of the four high/low valence/arousal quadrants. 

Many sources can elicit emotion. For instance, video, or lyrics may also affect our perceived or induced emotions. Some datasets offer alternative multimedia streams such as emotion-rated music videos from a variety of genres, including pop, rock, classical, and jazz as in the DEAP dataset \cite{koelstra2011deap}; or text of lyrics with the annotated music in the DMDD dataset \cite{delbouys2018music}. The MuVi dataset \cite{chua2022predicting} even offers isolated modality ratings for music videos. In this study, the raters were presented with either the music video, the music alone, or the muted video. The final dataset contains ratings for each of these modalities separately as well as together. This allows \cite{chua2022predicting} to build a model on pure isolated modalities, which proved to be more accurate than a traditional model. 

The RAVDESS dataset \cite{livingstone2018ryerson} is a multimodal dataset containing both speech and music with emotional expressions: calm, happy, sad, angry, fearful, surprise, and disgust for speech; and calm, happy, sad, angry, and fearful for music. It provides a rich combination of modalities, including Audio-only, Audio-Video, and Video-only formats.

Additionally, the EmoMV dataset \cite{thao2023emomv} focuses on affective music-video correspondence learning by providing labeled pairs of music and video segments that either match or mismatch in terms of emotional content. This dataset enables models to learn the affective alignment between audio and visual modalities, making it particularly valuable for emotion-based matching and retrieval tasks.

Recently, the Popular Hooks dataset \cite{wu2024popular} was introduced as a multimodal dataset that contains 38,694 popular musical hooks (i.e., memorable sections of songs) with synchronized MIDI, music video, audio, and lyrics. It also offers detailed (predicted) labels for high-level musical attributes such as tonality, structure, genre, emotion, and region. Leveraging a pre-trained multimodal music emotion recognition framework, the dataset provides predicted emotion labels, which were evaluated through a user study. 

Lastly, datasets like MERP \cite{koh2022merp} and EMOPIA \cite{hung2021emopia} provide metadata on song features and listener demographics, which can be useful for detailed analysis and personalized emotion prediction. 

\textbf{Dataset size}
Compared to affective datasets available in other domains (e.g., the Sentiment140 dataset \cite{go2009twitter} in NLP, which contains 1.6 million tweets annotated as positive or negative), the size of the available datasets is still very limited, with the largest dataset containing 109,269 instances. In addition to the number of instances in datasets, we also notice a difference in the length of the instances. A number of datasets (e.g. MERP \cite{koh2022merp}, VGMIDI \cite{ferreira2021learning}, and HKU956 \cite{hu2022detecting}) offer ratings for full-length songs. In datasets with dynamic ratings throughout the song, this may provide a means for researchers to analyse how our emotions evolve throughout a song. Other datasets focus on short fragments, often ranging from 30 seconds to 1-minute fragments (e.g. DEAM \cite{aljanaki2017developing}, EMOPIA \cite{hung2021emopia}, and EMMA \cite{strauss2024emotion}), but even as short as 10s as is the case for the MER500 dataset which consists of Indian Hindi film music \cite{velankar2020}. 

\textbf{Variety of music}
The genres covered in these datasets vary widely, reflecting the diverse nature of music. For instance, the DEAP dataset \cite{koelstra2011deap} includes classical as well as jazz pieces, which are known for their rich emotional and structural complexities. The VGMIDI dataset on the other hand, \cite{ferreira2021learning} focuses exclusively on video game soundtracks, a genre that often aims to evoke specific emotions to enhance the gaming experience. Rock and electronic music can primarily be found in the DEAM dataset \cite{aljanaki2017developing}, both genres that are prevalent in contemporary music culture. This genre diversity allows researchers to select datasets that align with their specific research goals and to explore how different genres affect emotional perception and annotation. Additionally, studying a variety of genres can help in developing more robust and generalized MER models that can perform well across different types of music. 

\textbf{Annotation Process}
Annotation processes and label distributions vary significantly across datasets, impacting the effectiveness and reliability of machine learning models in MER. Some datasets, such as MTG-Jamendo \cite{bogdanov2019mtg}, rely on freely assigned tags by multiple annotators, allowing a broad range of emotional descriptors. In contrast, DEAP \cite{koelstra2011deap} and HKU956 \cite{hu2022detecting} use a predetermined list of emotions to ensure consistency across annotations.

A major distinction between annotation strategies relates to if they use absolute or relative annotation methods, this is particularly important to note for dimensional models of emotion.
In absolute annotation, annotators assign a direct score or category to each sample independently. This approach is common in datasets such as DEAP and HKU956, where participants provide valence and arousal ratings individually for each stimulus. 
Recenlty, several datasets propose \textit{relative} annotation strategies to simplify the annotation process and improve consistency across annotators. Instead of rating individual tracks absolutely, annotators are asked to compare pairs of samples and judge their relative emotional positioning. Examples include the MusAV dataset \cite{bogdanov2022musav}, which gathers comparative annotations of arousal and valence for pairs of music tracks, and Emo-Soundscapes \cite{fan2017emo}, where participants rank soundscape recordings based on perceived emotion.
Earlier work by Yang et al. \cite{yang2010ranking} introduced ranking-based emotion recognition methods for music organization and retrieval, showing that relative judgments can lower cognitive load and lead to more reliable ground truth data. Similarly, in the EMusic dataset \cite{fan2017ranking}, a ranking-based annotation approach was used for experimental music, demonstrating improvements over traditional absolute labeling methods.

In addition, the CCMED-WCMED dataset \cite{fan2020comparative} compared Western and Chinese classical music based on emotion dimensions collected through relative annotations, highlighting the importance of cultural factors in emotion perception.
These relative approaches can mitigate some of the variability and subjectivity inherent in absolute annotations, particularly in continuous emotion models like valence-arousal space.

The number of annotators also varies: some datasets involve large-scale crowdsourcing efforts, providing a wide range of perspectives (e.g., MusAV, Emo-Soundscapes), while others use smaller, more controlled groups to maintain annotation consistency (e.g., DEAP, HKU956).
Additionally, the logic behind selecting music segments differs, with some datasets focusing on full-length songs and others on shorter clips to better capture transient emotional responses.

Overall, the choice between absolute and relative annotation methods, the number of annotators, and the segment selection strategies are all critical factors influencing dataset reliability and the resulting model performance in MER research.

\begin{figure*}[!t]
\centering
\includegraphics[width=6.8in]{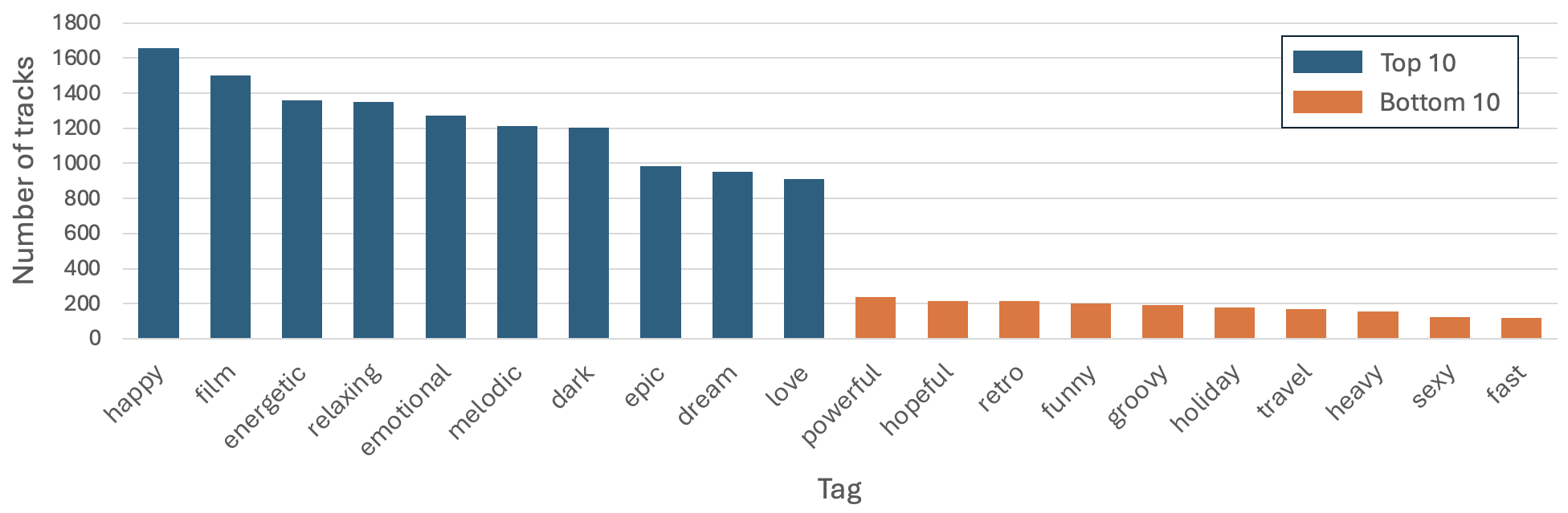}
\caption{Top 10 most and least frequent tags in the MTG-Jamendo dataset.}
\label{fig:dataset_frequency}
\end{figure*}

\textbf{Label distribution and splits}
The distribution of labels across datasets can be imbalanced, affecting the performance of machine learning models. For instance, the MTG-Jamendo dataset \cite{bogdanov2019mtg} contains a diverse set of 56 tags, freely assigned by users, but their frequency is highly imbalanced. This skewed distribution, with some tags like ``happy'' and ``film'' appearing far more often than others like ``sexy'' or ``fast'', is visualized in Figure~\ref{fig:dataset_frequency}, which shows the top 10 most and least frequent tags.
In contrast, the DEAP dataset \cite{koelstra2011deap} provides a more balanced set of labels across arousal and valence dimensions. 
Finally, some datasets, such as DEAM \cite{aljanaki2017developing} and MTG-Jamendo \cite{bogdanov2019mtg}, propose fixed splits for training and testing. This facilitates the replication of experiments and ensures consistent evaluation metrics which are easily used as a benchmark.

\textbf{Recommendations and Future Directions}
While the landscape of emotion-annotated music datasets is expanding, several challenges and opportunities remain. For general-purpose MER models, datasets like MTG-Jamendo \cite{bogdanov2019mtg} and MERP \cite{koh2022merp} offer breadth in genre and label diversity, while EMOPIA \cite{hung2021emopia} and VGMIDI \cite{ferreira2021learning} are more suitable for symbolic-domain or piano-focused emotion modeling. For multimodal applications, Popular Hooks \cite{wu2024popular} and EmoMV \cite{thao2023emomv} are relevant. However, there is a notable lack of large-scale, publicly available datasets focusing on induced emotions with multimodal or biosensor data, limiting medical and affective computing applications. The community would benefit from more datasets capturing diverse demographics, cross-cultural emotional interpretations, and consistent emotion annotation schemas across modalities. We encourage future dataset creators to include dynamic annotations, listener metadata, and align modalities when possible.

In sum, there is a variety of emotion-annotated datasets available as shown in Table~\ref{tab:datasets}. They differ in terms of emotion representation models, annotations labels, as well as music format and perceived versus induced emotion labels. Understanding these variations is crucial for developing robust and accurate MER models that can generalize across different datasets and musical genres. The challenges and future research directions related to datasets are discussed in detail in Section~\ref{sec:challenges}. 

\section{Evaluation Protocols}
\label{sec:eval}
Evaluation metrics play a crucial role in assessing the performance of MER systems. Depending on which type of emotion ratings are being predicted (dimensional versus categorical), the evaluation metrics change as we are dealing with a regression or a classification task respectively. Commonly used evaluation metrics for categorical MER systems include accuracy, precision, area under the ROC curve (AUC), and confusion matrices. In the case of regression MER models, metrics such as mean squared error (MSE), \( R^2 \), and Pearson correlation coefficient \cite{cohen2009pearson} are used. These metrics provide insights into the effectiveness of MER models when it comes to emotions represented by dimensional models.

The evaluation approach also depends on whether a dataset provides static or dynamic annotations. Static annotations assign a single label (or vector) to an entire track or segment, making evaluation straightforward with standard classification or regression metrics (e.g., accuracy, F1-score, ROC-AUC, or \( R^2 \)) computed at the song level \cite{bogdanov2019mtg, panda2018musical, yang2008regression}. In contrast, dynamic annotations provide time-continuous emotion ratings—often sampled at a rate of 1 Hz—reflecting how emotions evolve throughout a track \cite{aljanaki2017developing, koelstra2011deap, zhang2018pmemo}. In such cases, evaluation typically involves comparing predicted and ground-truth time series using frame-level metrics such as Pearson Correlation Coefficient (PCC), Mean Squared Error (MSE), or Concordance Correlation Coefficient (CCC) \cite{aljanaki2017developing, cheuk2020regression, zhang2018pmemo}. Some studies apply smoothing or window-based aggregation (e.g., moving average or median filters) to reduce annotation noise and capture broader emotional trends \cite{cheuk2020regression, aljanaki2017developing}. For example, the DEAM dataset \cite{aljanaki2017developing} provides both static and dynamic valence-arousal annotations, with dynamic evaluation conducted on a per-frame basis across time.

When evaluating MER models, it is also important to distinguish between perceived and induced (invoked) emotion. Evaluating perceived emotion predictions typically involves comparing predicted labels with human-annotated ground truth, which reflects how listeners interpret the emotional content of the music. In contrast, evaluating induced emotion is more complex, as it involves the actual emotional response elicited in the listener. This often requires physiological measurements (e.g., heart rate, EEG) or self-reports in a controlled experimental setup. For this reason, evaluation of induced emotion prediction models may rely on additional modalities (e.g., biosensors), and often focuses on subjective measures such as emotion regulation effectiveness or user satisfaction. These differences highlight the need for tailored evaluation protocols depending on the target emotion type.

Evaluating MER systems goes beyond just defining a common metric. Due to the inherent differences between datasets, a comparison across datasets is often not possible. For an in-depth discussion on this, the reader is referred to Section~\ref{sec:challenges}. Within one dataset, however, it is possible to establish benchmarks and compare the performance of different models, provided that the train/test split is shared. There are some initiatives to facilitate the comparison between models, such as competitions, as well as individual papers that offer clear data splits and metrics \cite{cheuk2020regression, rajamani2020emotion, mayerl2021recognizing,tan2021semi,bour2021frequency}.

Competitions and challenges provide valuable platforms for evaluating and comparing the performance of MER systems. These initiatives often involve standardized datasets, evaluation protocols, and metrics, enabling researchers to benchmark their algorithms against state-of-the-art methods. Existing benchmarking initiatives and competitions in MER include the `Audio K-POP Mood Classification' task in MIREX (Music Information Retrieval Evaluation eXchange) (last organized in 2019)\footnote{\url{https://www.music-ir.org/mirex/wiki/2019:Audio_K-POP_Mood_Classification/}}
, the `Emotion in Music' task in MediaEval (last organized in 2015)\footnote{\url{http://www.multimediaeval.org/mediaeval2015/emotioninmusic2015/}}, and the `Emotion and Theme Recognition in Music Using Jamendo' in MediaEval (last organized in 2021)\footnote{\url{https://multimediaeval.github.io/2021-Emotion-and-Theme-Recognition\\-in-Music-Task/}}.


Given the variability and potential noisiness of emotion annotations in music datasets~\cite{koh2022merp}, evaluation protocols should be designed to assess not only performance against possibly imperfect ground truth but also the perceptual validity of model outputs. One way to address this is by incorporating listening tests, where human subjects rate the emotions conveyed by model-predicted tracks. This does not directly mitigate the noise in the training data, but rather provides an additional, human-centered validation of the model’s effectiveness. Another complementary evaluation strategy is extrinsic evaluation, where emotion recognition is applied in downstream tasks. For instance, predicted emotions could be used to generate emotionally coherent playlists, and users could be asked whether the playlists evoke the intended emotional responses. Such evaluations help assess whether MER systems produce outputs that are meaningful and useful in real-world applications, beyond what can be inferred from standard metrics alone.

\begin{table*}[htp]
\small
\centering
\caption{Overview of Selected Music Emotion Recognition Models since 2020. Note: V = Valence, A = Arousal, Q = Quadrant.  }
\label{tab:models}
\resizebox{\textwidth}{!}{ 
\begin{tabular}{@{}lllp{5.8cm}p{2.7cm}p{2cm}p{5.1cm}@{}}
\toprule
\textbf{Ref.} & \textbf{Year} & \textbf{Modalities} & \textbf{Approach} & \textbf{Emotion Model} & \textbf{Dataset} & \textbf{Performance} \\
\midrule
\cite{sharma2020new} & 2020 & Audio, Lyrics & Machine Learning (e.g., SVM, NB) & Russell & PMEmo & Accuracy: 63\% \\

\cite{he2020multi} & 2020 & Audio & CNN, LSTM & Russell & Solymani et al.’s dataset & RMSE: 0.219 V, 0.212 A  \\

\cite{rajesh2020musical} & 2020 & Audio & LSTM & happy, sad, neutral, fear & Self-built & 
Accuracy: 89.3\%\\

\cite{chen2020multimodal} & 2020 & Audio, Lyrics & CNN-LSTM & angry, happy, relaxed, sad & Last.fm & Accuracy: 78\% \\

\cite{chaki2020attentive} & 2020 & Audio & Attentive LSTM & Russell & Solymani et al.’s dataset & R2: 0.53 V, 
 0.75 A 
 \\

\cite{de2020multiple} & 2020 & Audio & Source Separation, CNN & Russell & PMEmo & R2: 0.4814 V, 0.6004 A \\

\cite{russo2020cochleogram} & 2020 & Audio & Cochleogram, CNN & Russell & Solymani et al.’s dataset & R2: 0.41 V, 0.63 A \\

\cite{sarkar2020recognition} & 2020 & Audio & CNN, Local Attention & Quadrants & Soundtrack, Bi-Modal & Accuracy (Soundtrack): 67.71\% \newline  F1 (Bi-Modal): 77.82\%\\

\cite{rajamani2020emotion} & 2020 & Audio & Attention-based Neural Networks & 56 mood/theme tags & MTG-Jamendo & PR-AUC: 0.118, ROC-AUC: 0.735\\

\cite{cheuk2020regression} & 2020 & Audio & Triplet Neural Networks & Russell &Solymani et al.’s dataset, DEAM& R2 (Solymani.): 0.378 V, 0.638 A \newline R2 (DEAM): 0.361 V, 0.672 A\\

\cite{knox2020mediaeval} & 2020 & Audio & CNN Ensemble, Focal Loss, Receptive Field Tuning & 56 mood/theme tags & MTG-Jamendo & PR-AUC: 0.161, ROC-AUC: 0.781 \\

\cite{yu2021research} & 2021 & Audio & CNN-LSTM & Russell & DEAM & Accuracy: 64.9\% \\

\cite{hizlisoy2021music} & 2021 & Audio & CNN, LSTM+DNN & 3 classes (high arousal + pos./neg. valence) & Self-built & Accuracy: 99.19\% \\

\cite{mayerl2021recognizing} & 2021 & Audio & Clustering-based Ensembles & 56 mood/theme tags & MTG-Jamendo & PR-AUC: 0.109, ROC-AUC: 0.705 \\

\cite{tan2021semi} & 2021 & Audio & Semi-Supervised, Noisy Student Training & 56 mood/theme tags & MTG-Jamendo & PR-AUC: 0.136, ROC-AUC: 0.769 \\

\cite{bour2021frequency} & 2021 & Audio & Frequency Dependent Convolutions & 56 mood/theme tags & MTG-Jamendo & PR-AUC: 0.151, ROC-AUC: 0.775 \\

\cite{pham2021selab} & 2021 & Audio & CNN, Co-teaching Training Strategy & 56 mood/theme tags & MTG-Jamendo & PR-AUC: 0.144, ROC-AUC: 0.760 \\

\cite{grekow2021music} & 2021 & Audio & LSTM, Pretrained Models & Russell & Self-built & R2: 0.46 V, 0.73 A \\

\cite{agrawal2021transformer} & 2021 & Lyrics & Transformers & Russell & MoodyLyrics, MER & Accuracy (MoodyLyrics): 94.78\% \newline Accuracy (MER): 94.44\% V,  
 88.89\% A, \newline 88.89\% Q\\

\cite{chowdhury2021tracing} & 2021 & Audio & Source-separation-based Explainer & Russell & DEAM, Midlevel, PMEmo & R2 (DEAM): 0.48 V, 0.50 A \newline R2 (PMEmo): 0.50 V, 0.65 A \\

\cite{huang2021generative} & 2021 & Audio & Generative Adversarial Network & happy, sad & Self-built & Accuracy: 87\% \\

\cite{pandeya2021deep} & 2021 & Audio, Video & CNN & exciting, fear, neutral, relaxation, sad, tension & Self-built & Accuracy: 88.56\% \\

\cite{griffiths2021multi} & 2021 & Audio & Linear Regressors & Russell & Self-built & R2: 0.78 V, 0.85 A \\

\cite{xia2022study} & 2022 & Audio & Clustering, Machine Learning (e.g., SVM) & Russell & Solymani et al.'s dataset & Accuracy: 84.9\% \\

\cite{qiu2022novel} & 2022 & MIDI & Multi-Task Learning & Quadrants & EMOPIA, VGMIDI & Accuracy (EMOPIA): 67.58\% \newline Accuracy (VGMIDI): 55.85\%  \\

\cite{cai2022feature} & 2022 & Audio & Feature Selection, SVR, RF & Russell & DEAM & R2: 0.587 V, 0.645 A \\

\cite{chua2022predicting} & 2022 & Audio, Video & LSTM & Russell & Self-built & RMSE: 0.1331 V, 0.0973 A \\

\cite{matos2022merge} & 2022 & Lyrics & Deep Learning, BERT & Quadrants & MIR Lyrics Emotion & F1: 88.9 \%\\

\cite{he2022music} & 2022 & Audio & CNN, LSTM & Russell & PMEmo, AllMusic & Accuracy (PMEmo): \newline 79.01\% V, 83.62\% A \newline Accuracy (AllMusic): \newline 67.11\% V, 86.56\% A \\

\cite{medina2022emotional} & 2022 & Audio & SVM, Random Forest, MLP & Russell & DEAM & RMSE: 0.23 V,  0.24 A \\

\cite{krols2023multi} & 2023 & Audio, Lyrics & Multi-Modality & Russell & DMDD & R2: 0.235 V, 
 0.196 A  \\

\cite{han2023music} & 2023 & Audio & Inception-GRU Residual & Quadrants & Soundtrack & Accuracy: 84\% \\

\cite{song2023modeling} & 2023 & Lyrics & State Space Models & anger, disgust, fear, joy, sadness, surprise & LyricsEmotions & Pearson correlation: 0.345 (angry), \newline 0.268 (disgust), 0.350 (fear), \newline 0.503 (joy), 0.350 (sad), \newline 0.089 (surprise)\\

\cite{shanker2023tollywood} & 2023 & Lyrics & Fine-tuned XLMRoBERTa & Russell & Self-built & Accuracy: 80.88\% V, 81.51\% A, \newline 62.58\% Q \\

\cite{sams2023multimodal} & 2023 & Audio, Lyrics & CNN-LSTM, XLNet Transformers & positive, neutral, negative & Self-built & Accuracy: 80.56\% \\

\cite{zhang2023modularized} & 2023 & Audio & Attention Mechanism & Russell & DEAM, PMEmo & RMSE (DEAM): 0.112 V, 0.109 A \newline RMSE (PMEmo): 0.144 V, 0.135 A\\

\cite{lucia2023automatic} & 2023 & Audio & CNN & happiness, fear, sadness, peacefulness & Musical Excerpts & Accuracy: 92\%\\

\cite{suresh2023transformer} & 2023 & Audio, Lyrics & Transformers & Quadrants & MoodyLyrics & Accuracy: 77.94\%\\

\cite{chang2024music} & 2024 & Audio & RNN, BRNN, LSTM & Quadrants & 4Q audio, MTG-Jamendo & Accuracy (4Q.): 65.97\% \newline Accuracy (MTG.): 53.97\% \\

\cite{wang2024mmd} & 2024 & Audio, Lyrics & VGGish, ALBERT & happy, sad, calm, healing & DEAM, FMA & Accuracy (DEAM): 49.68\% \newline  Accuracy (FMA): 49.54\% \\

\bottomrule
\end{tabular}
}
\end{table*}

\addtocounter{table}{-1}
\begin{table*}[htp]
\small
\centering

\caption{(continued)}

\label{tab:models}
\resizebox{\textwidth}{!}{ 
\begin{tabular}{@{}lllp{6cm}p{2.5cm}p{2cm}p{5.1cm}@{}}
\toprule
\textbf{Ref.} & \textbf{Year} & \textbf{Modalities} & \textbf{Approach} & \textbf{Emotion Model} & \textbf{Dataset} & \textbf{Performance} \\
\midrule

\cite{raboy2024verse1} & 2024 & Audio, Lyrics & Stacked Ensemble Models & angry, happy, relax, sad & Self-built & Accuracy: 96.25\%\\

\cite{han2024gai} & 2024 & Audio & Multi-scale Parallel Convolution & sleepy, calm, sad, pleased, relaxed, nervous, annoying, excited & PMEmo, Soundtrack, RAVDESS & Accuracy: 98.58\%\\

\cite{li2024improved} & 2024 & Audio & CNN with Differential Evolution & happy, sad, angry, calm & self-built, DEAM & Accuracy: 85.29\% \\

\cite{liu2024leveraging} & 2024 & Audio & LLM Embeddings, Non-parametric Clustering & Categorial & MTG-Jamendo, CAL500, Emotify & F1 (MTG-Jamendo): 2.02\% \newline F1 (CAL500): 22.9\% \newline F1 (Emotify): 40.0\% \\

\cite{kang2025towards} & 2025 & Audio & MERT Embeddings, Multitask Learning, Transformers, Knowledge Distillation & Categorial, Russell & MTG-Jamendo, PMEmo, DEAM, Solymani et al.’s dataset & (MTG-Jamendo) PR-AUC: 0.1543 \newline ROC-AUC: 0.7810 \newline (Solymani.) R2: 0.6512 V, 0.7616 A \newline (PMEmo) R2: 0.5473 V, 0.7940 A \newline (DEAM) R2: 0.5184 V, 0.6228 A
\\

\bottomrule
\end{tabular}
}
\end{table*}

\section{Models and approaches}
\label{sec:5}
In this section, we briefly touch upon some of the more recent MER models, highlighting the state-of-the-art approaches over the last few years. This is not an exhaustive overview, and for a more comprehensive review, readers are referred to other survey papers \cite{han2022survey, kim2010music, yang2018review, yang2012machine, barthet2013music}. The aim of this section is to point out the \emph{current state-of-the-art} and various approaches over the last five years. The models were selected through Google Scholar searches using the search terms `music emotion prediction model', `affective music prediction model', and `music emotion recognition model' starting from the year 2020, as well as by following links in the articles. 

Table \ref{tab:models} presents an overview of selected MER models released since 2020, summarizing their modalities, approaches, emotion models, and datasets. This table highlights the diversity in methodologies and datasets used in recent research.

Some of the earliest attempts at music emotion prediction involved rule-based approaches and hierarchical frameworks. For instance, Feng et al. \cite{feng2003music} used Computational Media Aesthetics (CMA) to analyze tempo and articulation, mapping them into four mood categories: happiness, anger, sadness, and fear. They achieved a total precision of 67\% and a total recall of 66\%. Lu et al. \cite{lu2005automatic} developed a hierarchical framework for automatically extracting music emotion from acoustic data. They employed music intensity to represent the energy dimension of Thayer's model while using timbre and rhythm to capture the stress dimension. They achieved an average accuracy of mood detection of up to 86.3\%. These early models laid the groundwork for subsequent advancements in music emotion recognition, paving the way for the adoption of more sophisticated techniques, including deep learning approaches.

\begin{figure*}[!t]
\centering
\includegraphics[width=6.8in]{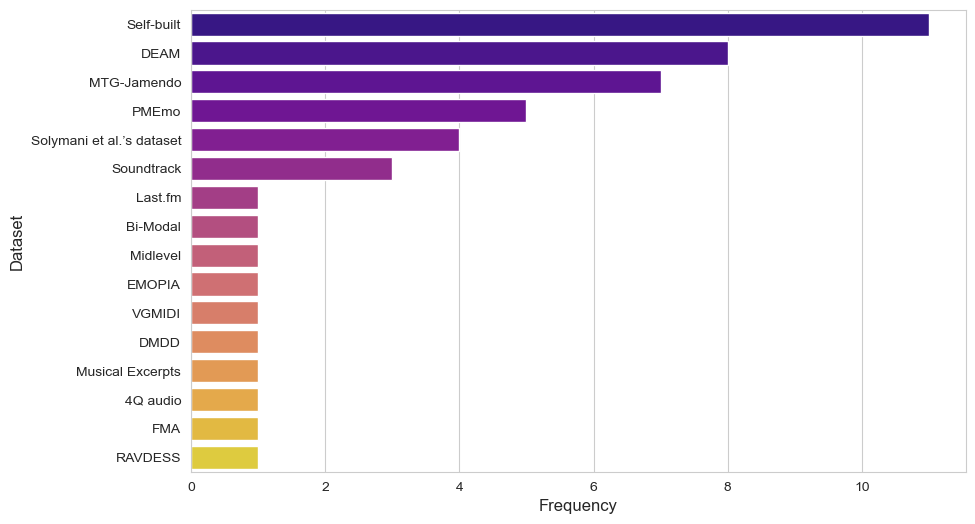}
\caption{Number of Music Emotion Recognition Models since 2020 that use the listed datasets.}
\label{fig:dataset_frequency}
\end{figure*}


In recent years, various deep-learning MER models have been developed, employing techniques such as Convolutional Neural Networks (CNNs) \cite{he2020multi, chen2020multimodal, de2020multiple, russo2020cochleogram, sarkar2020recognition, rajamani2020emotion, pandeya2021deep, yu2021research, hizlisoy2021music, pham2021selab, he2022music, lucia2023automatic, sams2023multimodal, han2023music, wang2024mmd, li2024improved, han2024gai}, Recursive Neural Networks (RNNs) such as Gated Recurrent Units (GRUs) \cite{han2023music}  and Long-Short Term Memory networks (LSTMs) \cite{he2020multi, rajesh2020musical, chen2020multimodal, chaki2020attentive, yu2021research, hizlisoy2021music, grekow2021music, he2022music, chua2022predicting, sams2023multimodal, chang2024music}, and more recently, Transformer architectures \cite{agrawal2021transformer, suresh2023transformer, sams2023multimodal, shanker2023tollywood}. 

Additionally, generative models such as Generative Adversarial Networks (GANs) have been applied. For instance, Huang et al. \cite{huang2021generative} developed a GAN-based model for emotion recognition using audio inputs under IoT environments.

On the MTG-Jamendo dataset \cite{bogdanov2019mtg}, several models have demonstrated noteworthy performance. For example, Mayerl et al. \cite{mayerl2021recognizing} employed a CNN-based (ResNet18) model using clustering-based ensembles, achieving a PR-AUC-macro score of 0.1087 and a ROC-AUC-macro score of 0.7047. Tan et al. \cite{tan2021semi}, using a semi-supervised learning method and a CRNN architecture, reported higher scores of PR-AUC-macro 0.1357 and ROC-AUC-macro 0.7687. Pham et al. \cite{pham2021selab} achieved similar results with an EfficientNet-based CNN model, utilizing a co-teaching strategy to manage noisy data, resulting in PR-AUC-macro 0.1435 and ROC-AUC-macro 0.7599. 
Bour et al.~\cite{bour2021frequency} leveraged frequency-dependent convolutions in a CNN model, achieving PR-AUC-macro 0.1509 and ROC-AUC-macro 0.7748. 
Kang and Herremans~\cite{kang2025towards} proposed a unified multitask learning framework that leverages both categorical and dimensional emotion labels by combining musical features (e.g., key and chords) with MERT representations. Through knowledge distillation across multiple datasets, this approach enhances cross-dataset generalization and achieves improved performance on MTG-Jamendo with a PR-AUC of 0.1543 and ROC-AUC of 0.7810. Knox et al. \cite{knox2020mediaeval} proposed an ensemble of CNN-based models trained on Mel spectrograms, exploring the impact of different loss functions and resampling strategies for multi-label music tagging. They showed that using focal loss effectively addressed class imbalance, and adjusting the CNN’s receptive field further improved performance. Their model achieved the highest reported scores in the MediaEval ``Emotion and Theme Recognition in Music'' task, with a PR-AUC-macro of 0.1610 and ROC-AUC-macro of 0.7810.


Notably, hybrid architectures combining multiple deep learning techniques, such as CNN-LSTM models, have shown promising results. For instance, Chen and Li \cite{chen2020multimodal} employed a CNN-LSTM model for emotion recognition using the Last.fm subset of the Million Song Dataset—a large-scale collection of music metadata with listener-generated tags. Their model outperformed standalone CNN and LSTM baselines, achieving an accuracy of 68.1\% for audio classification and 74.2\% for lyric classification. In contrast to classification approaches, He et al.~\cite{he2020multi} proposed a regression-based model using multi-view CNNs and bidirectional LSTMs directly on raw audio. Without relying on handcrafted features, their model achieved strong performance with RMSE scores of 0.219 for valence and 0.212 for arousal on the Solymani et al.’s dataset \cite{soleymani20131000}. Transformer-based architectures have also gained traction in recent years, with notable examples by Suresh et al. \cite{suresh2023transformer} and Agrawal et al. \cite{agrawal2021transformer}, who explored their effectiveness in MER. Suresh et al., for example, applied a Transformer model to the MoodyLyrics dataset \cite{ccano2017moodylyrics}, which contains 2,595 tracks labeled with four mood categories derived from lyrics. Their model achieved the highest reported accuracy of 77.94\%, outperforming CNN and Bi-GRU baselines.


Other models have employed more traditional machine learning techniques alongside deep learning approaches.
Medina et al. \cite{medina2022emotional} focused on classifying emotions from audio using SVM, Random Forest, and Multi-Layer Perceptron (MLP) models, achieving an F-score of 0.73 and 0.69 for predicting valence and arousal values, respectively, on the DEAM dataset \cite{aljanaki2017developing}. Sharma et al. \cite{sharma2020new} combined machine learning algorithms such as Support Vector Machines (SVM) and Naive Bayes (NB) to predict emotions using both audio and lyric features. They achieved an accuracy of 63\% on the PMEmo dataset \cite{zhang2018pmemo}. Similarly, Griffiths et al. \cite{griffiths2021multi} developed a multi-genre MER model using linear regressors. They achieved a \( R^2 \) score of 0.776 and 0.85 for predicting valence and arousal values, respectively, on the self-built dataset. Meanwhile, Xia et al. \cite{xia2022study} utilized clustering algorithms alongside machine learning techniques like SVM and K-Nearest Neighbors (KNN) for emotion recognition, and their hybrid model which combined all the machine learning techniques achieved an accuracy of 85\% on the Solymani et al.'s dataset \cite{soleymani20131000}.

A distinction between the various models can be made based on the input modality. Some models are exclusively based on MIDI such as the multi-task architecture proposed by \cite{qiu2022novel}, and as such use a token-based representation. Given the limited size of MIDI datasets, the accuracy of such models is limited, e.g., the model by \cite{qiu2022novel} reaches 67.56\% accuracy when predicting between four emotion classes. 
Most of the existing MER models are based on audio, and hence they take as input raw audio. This is often converted into Mel-spectrograms which are then processed through convolutional neural networks (CNNs) \cite{sarkar2020recognition}, or the audio could be directly processed through a WaveNet architecture, which is a type of temporal CNN \cite{do2020hcmus}. 

In recent years, audio embeddings pretrained on large-scale datasets have become increasingly common, enabling transfer learning for Music Emotion Recognition (MER). Beyond traditional embeddings, recent research has explored both supervised and unsupervised pretraining strategies for audio representation learning, where MER is treated as a downstream task. For instance, MERT (Music undERstanding model with self-supervised Training) \cite{li2023mert} is specifically tailored to music, addressing challenges like pitch and tonality through a novel training scheme. MERT incorporates pseudo labels from two teacher models—an acoustic teacher based on Residual Vector Quantization Variational AutoEncoder (RVQ-VAE) and a musical teacher based on the Constant-Q Transform (CQT)—within a masked language modeling framework. When used as a feature extractor for MER, MERT-95M achieves a PR-AUC of 0.1340 and ROC-AUC of 0.7640 on the MTG-Jamendo dataset \cite{bogdanov2019mtg}, while MERT-330M reaches a PR-AUC of 0.1400 and ROC-AUC of 0.7650. Building upon MERT embeddings, Kang and Herremans~\cite{kang2025towards} introduce a unified multitask learning framework that leverages both categorical and dimensional emotion labels. Their model achieved a PR-AUC of 0.1543 and ROC-AUC of 0.7810 on the MTG-Jamendo dataset \cite{bogdanov2019mtg}. For dimensional emotion regression, it attained \( R^2 \) scores of 0.5473 for valence and 0.7940 for arousal on PMEmo \cite{zhang2018pmemo}, 0.5184 for valence and 0.6228 for arousal on DEAM \cite{aljanaki2017developing}, and 0.6512 for valence and 0.7616 for arousal on the dataset by Solymani et al. \cite{soleymani20131000}.


McCallum et al. \cite{mccallum2022supervised} present a comparative analysis of audio representation learning strategies, showing that supervised pretraining on expert-annotated music datasets leads to state-of-the-art performance in emotion tagging. Their models achieve 78.6 ROC-AUC and 16.1 PR-AUC on the MTG-Jamendo Mood/Theme dataset—outperforming previously reported benchmarks. Alonso-Jiménez et al. \cite{alonso2023efficient} propose MAEST, a transformer-based architecture trained with patchout and pre-initialized with ImageNet or AudioSet weights. Although its performance is slightly lower (78.1 ROC-AUC and 15.4 PR-AUC), MAEST demonstrates the potential of efficient, convolution-free architectures in music emotion tagging, particularly when speed and scalability are important.


Alternatives to these approaches include directly extracting spectral features (e.g., MFCCs, spectral centroids) with libraries such as OpenSmile, which has a configuration file specifically for the emotion recognition task \cite{eyben2010opensmile}. \cite{cheuk2020regression} uses this approach and achieves \( R^2 \) scores of 0.378 and 0.638 for predicting dynamic valence and arousal values, respectively, on the Solymani et al.'s dataset \cite{soleymani20131000}. Recognizing the importance of induced emotions, recent research has focused on models that leverage physiological data to predict music-induced emotions. For instance, \cite{hu2022detecting} constructed the HKU956 dataset \cite{hu2022detecting} with aligned peripheral physiological signals (i.e., heart rate, skin conductance, blood volume pulse, skin temperature) and self-reported emotion from 30 participants. The study revealed that physiological features significantly contribute to valence classification and that multimodal classifiers outperform single-modality ones. \cite{chowdhury2021tracing} uses PMEmo \cite{zhang2018pmemo} as induced emotion recognition dataset. This study merges audioLIME, a source-separation-based explainable model, with mid-level perceptual features to form an intuitive connection between input audio and emotion predictions, providing insights into model predictions. \cite{zhang2023modularized} also uses PMEmo as induced emotion recognition dataset and introduces a novel method named Modularized Composite Attention Network (MCAN). This method enhances feature extraction and employs attention mechanisms to improve the stability and accuracy of emotion prediction models.

Finally, musically meaningful features may be included, such as Rhythmic features (e.g., tempo, beat histogram), or note features (e.g., pitch), as implemented by Shi et al. \cite{shi2006tempo} and Panda et al. \cite{panda2018musical}, respectively. Shi et al. achieved a precision of 92.8\% for 4 emotion categories (calm, sad, pleasant, and excited), while Panda et al. achieved an F1-score of 76.0\% for 4 emotion categories (Quadrants).

Other input modalities may include text (lyrics), or video. In the case of the former, some models extract the sentiment from the lyrics using Natural Language Processing (NLP) tools \cite{krols2023multi, matos2022merge}, or they use the entire lyrics with an embedding model \cite{suresh2023transformer, song2023modeling}. These features are then combined with audio-based features or analyzed independently to predict the emotional content of music. Including the lyrics does not always improve the sentiment prediction, particularly in predicting arousal \cite{krols2023multi}; however, \cite{raboy2024verse1} did manage to increase the performance of a MER model by using embeddings, such as word2vec \cite{mikolov2013efficient} or stacked ensemble models that integrate both audio and lyrics. Finally, models that include video modalities typically use pretrained networks to capture image and video features, and thus improve model performance. For instance, \cite{thao2023emomv} uses ResNet-50 \cite{he2016deep} and FlowNetS \cite{dosovitskiy2015flownet}, respectively.

Identifying the current state-of-the-art Music Emotion Recognition (MER) model is challenging due to factors such as differences in datasets and performance metrics, as highlighted in the previous section. 
A commonly used benchmark is the Emotion and Theme Recognition in Music competition based on the MTG-Jamendo dataset, where the top-performing model by Knox et al. \cite{knox2020mediaeval} achieved a PR-AUC-macro of 0.161 and a ROC-AUC-macro of 0.781 using an ensemble of CNN-based models trained with focal loss and receptive field tuning. While benchmark performances of MER models have steadily improved (e.g., some achieving ROC-AUC scores over 0.78 on datasets like MTG-Jamendo), their translation to real-world applications remains limited.

One area where we do see emotion models integrated is emotion-conditioned music generation, where preliminary MER models are incorporated to guide the emotional content of generated music (e.g., Makris et al.~\cite{makris2021generating}). Furthermore, emotion detection systems have been piloted in adaptive gaming environments or personalized music therapy applications \cite{agres2021music}, but large-scale real-world deployment of deep MER models remains rare due to variability across listeners, subjective annotation challenges, and differences between training and inference conditions.



Figure~\ref{fig:dataset_frequency} illustrates the most frequently used datasets by the listed Music Emotion Recognition (MER) models since 2020, as summarized in Table~\ref{tab:models}. This count includes studies where MER is the primary task as well as those where it serves as a downstream application, such as in music representation learning \cite{li2023mert}. Self-built datasets, created by authors for specific research purposes, remain the most commonly used, appearing in 11 instances. These datasets often contain licensed music, restricting their accessibility to the broader research community. Other frequently utilized datasets include DEAM (8 occurrences) and MTG-Jamendo (7 occurrences), highlighting their importance in MER studies. Some datasets, such as PMEmo \cite{zhang2018pmemo} and Solymani et al.'s dataset \cite{soleymani20131000}, appear less frequently but still contribute valuable insights. For instance, PMEmo \cite{zhang2018pmemo} includes dynamic emotion labels and physiological signals (EDA), enabling multimodal affective analysis, while Solymani et al.’s dataset \cite{soleymani20131000} offers continuous valence-arousal annotations and standard deviations, supporting studies on annotation reliability and temporal emotion dynamics.

We have only provided a glimpse into the existing MER models in this section. From the performance of the various models, we see, however, that much improvement can still be made. We discuss some of the remaining challenges in the next section. 

\section{Challenges and future directions}
\label{sec:6}
\label{sec:challenges}
Whereas the first publications on music and emotion surfaced in the 1930s \cite{hevner1936experimental}, current MER models still struggle to match human performance in emotion recognition. The field of MER still faces several challenges and various opportunities for future exploration remain, ranging from overcoming data limitations to the integration of emerging technologies. Understanding and addressing these challenges is crucial for advancing the field.

\textbf{Dataset limitations} One of the primary challenges in MER is the scarcity of large, diverse, copyright-cleared, emotion-annotated datasets. Limited datasets hinder the development and evaluation of robust MER models, leading to potential biases and generalization issues, and also prevent the establishment of reliable performance benchmarks across studies.

We have recently seen advances in this area with the release of larger datasets such as MTG-Jamendo \cite{bogdanov2019mtg}, Music4all \cite{santana2020music4all}, MuSe \cite{akiki2021muse}, which contain 18k, 109k, and 90k instances respectively. Whereas MTG-Jamendo and MuSe are available under a Creative Commons licence, however, Music4all contains copyrighted tracks. 
In addition, when exploring datasets, we notice that they are often skewed towards one particular genre. For instance, the DEAM dataset \cite{aljanaki2017developing} consists mostly of rock and electronic music genres, while the VGMIDI dataset is focused solely on video game soundtracks.

To deal with the current dataset size limitations, techniques such as unsupervised pretrained may be helpful. With this technique, latent representations are first learned using unlabeled datasets. This evolution has led to some available large-scale audio encoders such as the Variational Auto Encoder used in \cite{melechovsky2023mustango}, and the AST Audio Spectrogram Transformer presented in \cite{gong2021ast}, as well as various recently developed neural audio encoders (e.g. Descript Audio Codec \cite{kumar2024high}). These novel pretrained representations may help deal with the limiting size of emotion-annotated datasets. 

\textbf{Subjective labels}
The subjective and variable nature of emotion perception also poses a significant challenge. Emotions are inherently complex and subjective, varying across individuals, cultures, and contexts \cite{koh2022merp}. For instance, researchers have observed significant dependence between the number of years of musical training \cite{lima2011emotion, koh2022merp}, gender \cite{chua2022predicting}, familiarity with songs \cite{chua2022predicting}, culture \cite{wang2021cross, koh2022merp}, genre preference \cite{koh2022merp}, and even age \cite{koh2022merp}. The latter study makes an argument for including profile information in the emotion prediction model to personalize the predictions. However, not many datasets include this type of information other than MERP. 

In addition, many datasets have a cultural bias, meaning that they are annotated by people from the same culture, often simply because of the locality of the experiment or language constraints. However, it has been established that people from different countries or cultural backgrounds have different perceptions of emotion for the same music fragment \cite{lee2021cross}. For instance, \cite{koh2022merp} have raters from both the US as well as India, and see a significant difference in their annotations. HKU956 \cite{hu2022detecting} go even further and include the rater's responses to a personality test: `Ten Item Personality Measure'. In future work, one could develop datasets annotated by annotators from different cultural backgrounds, that include this information about the raters. 
The personalized nature of emotion ratings can cause a low inter-rater reliability, a metric of agreement between raters often calculated using Cronbach's Alpha \cite{bland1997statistics}. 

\textbf{Noisy labels} 
When creating datasets, an additional challenge arises: it can be hard to identify the emotion perceived from music, especially when working with valence and arousal models. In fact, Russell's model \cite{russell1980circumplex} included a third dimension: dominance. This dimension is typically omitted because of the ambiguity in annotation. In general, categorical models, although less precise, are often easier to annotate \cite{song2016perceived}. This personal variability and ambiguity in emotion labels introduces noise in dataset labels, causing low inter-rater reliability, and making it challenging to train accurate and reliable MER models. In addition, some emotion representations are prone to more noise. For instance, the MTG-Jamendo dataset \cite{bogdanov2019mtg} offers a large number of freely assigned tags, making it suitable for exploring a wide range of emotions, but the lack of a controlled annotation process can lead to noisy data, as there may be synonyms of emotion terms. 

Machine learning models may also offer useful techniques to deal with noisy labels, as discussed in the survey by \cite{song2022learning}. These could include Noisy Graph Cleaning (NGC) \cite{wu2021ngc}, Joint Training with Co-Regularization (JoCoR) \cite{wei2020combating}, and Robust Curriculum Learning (RoCL) \cite{zhou2020robust}. 

\textbf{Annotation interfaces}
When creating a static dataset with static annotations, some standard tools can be used, including PsyToolkit \cite{stoet2017psytoolkit}. However, to create any larger-scale dataset, the annotations are often done through crowdsourcing services such as Amazon Mechanical Turk\footnote{\url{https://www.mturk.com}} (e.g. for MERP  \cite{koh2022merp}, DEAM \cite{aljanaki2017developing}, and Solyman’s dataset \cite{soleymani20131000}). These services offer access to a large `army' of annotators, which may come at the cost of accuracy. There are, however, a number of techniques that can be used to filter out some of this annotation noise. This includes limiting the annotations to Master raters (raters with a known track record that often work at a premium price) \cite{koh2022merp}, by including qualification tasks to assess participants' understanding of the dimensional model \cite{soleymani20131000}, or by using multiple ground truth questions that are shown to all raters \cite{koh2022merp}. Finally, inter-rater reliability can be used to filter out low-quality annotations \cite{koh2022merp}. When doing this, it is important to keep in mind personal characteristics, which may cause different people to rate music differently. Hence, inter-rater reliability should ideally be calculated by taking into account the rater's profile features. 

An additional difficulty is that the amount of available interfaces for music emotion annotation is very limited. Especially when it comes to time-continuous annotations of valence and arousal. \cite{koh2022merp} released their dynamic annotation interface\footnote{\url{https://github.com/dorienh/MERP}} that hooks into Amazon mTurk.  Kim et al. \cite{kim2008moodswings} also introduced, an annotation interface called MoodSwings, designed to record dynamic emotion labels.

\textbf{Benchmarking}
The ImageNet competition has been instrumental in establishing a clear performance benchmark among computer vision models \cite{deng2009imagenet}. While there have been similar competitions for music emotion prediction (see Section~\ref{sec:eval}), none of these competitions ran in the last three years, indicating the lack of a current benchmark for MER systems. To establish benchmarks on individual datasets, we have to revert to individual model papers, such as \cite{aljanaki2017developing, soleymani20131000}. It is unfortunately not always clear which train/test split these systems use, making direct comparisons hard. One way to facilitate an easy overview would be to leverage Leaderboard features on popular websites such as Papers With 
Code\footnote{\url{https://paperswithcode.com/sota}} and HuggingFace\footnote{\url{https://huggingface.co/docs/competitions/en/leaderboard}}. 

While there have been some efforts toward cross-dataset comparisons, such practices are still relatively rare. A notable example is the MusAV dataset \cite{bogdanov2022musav}, which provides a benchmark for evaluating arousal-valence (AV) regression models trained on different datasets. MusAV uses relative pairwise comparisons as ground truth and enables comparative validation across models trained on diverse AV datasets. Such initiatives are promising steps toward better generalization and standardized evaluation, but they are currently exceptions rather than the rule. In general, many datasets still use different emotion representations, and cross-dataset evaluation remains challenging. Recent efforts have aimed to address this limitation: Kang et al. \cite{kang2025towards} propose a unified multitask framework to combine categorical and dimensional labels, while Liu et al. \cite{liu2024leveraging} use LLM-based label embeddings to align emotion annotations across datasets and enable zero-shot generalization.

Bridging between datasets that use continuous representations and categorical models is not straightforward. Several studies, such as Paltoglou and Thelwall \cite{paltoglou2012seeing}, have proposed mappings between arousal/valence and categorical labels. This approach has been further utilized, for example, by Makris et al. \cite{makris2021generating} for emotion-controlled lead sheet generation. In addition, large-scale affective norm resources such as Warriner et al. \cite{warriner2013norms}, which provides valence, arousal, and dominance (VAD) ratings for 13,915 English lemmas, and Mohammad \cite{mohammad2018obtaining}, who developed the NRC VAD Lexicon with ratings for over 20,000 words, offer valuable foundations for linking continuous and categorical representations. Such mappings could support merging continuous and categorical datasets into larger-scale resources for music emotion research.

\textbf{MIDI} The forgotten format in music emotion recognition. Emotion originates from many different aspects of the music, including the tonal tension \cite{herremans2017morpheus}, instrumentation and timbre \cite{mcadams2014perception}, production quality \cite{ronan2018empirical}, expressiveness of the performance \cite{juslin2010expression}, harmony \cite{farbood2006quantitative}. The symbolic MIDI format only captures parts of these \cite{kim2010music}, which may explain the lack of models for predicting emotion from MIDI. Another reason may simply be the lack of emotion-annotated datasets (three in total). Even though MIDI is an abstraction of music, it is still a widely used format by music producers and performers and warrants its own models for emotion prediction. Even more, such datasets may enable generative music systems (which are typically trained on MIDI) to be controlled by emotion \cite{makris2021generating, tan2020music}. 

\textbf{Multimodal predictions}
Our sensory input is multimodal. As such, some of the emotion-annotated datasets enable us to look at multiple modalities. For instance, DEAP \cite{koelstra2011deap} offers biofeedback data, i.e. EEG recordings, and frontal face videos from participants. Similarly, HKU956 \cite{hu2022detecting} offers physiological signals including heart rate, electrodermal activity, blood volume pulse, inter-beat interval, and skin temperature. SiTunes \cite{grigorev2024situnes}, on the other hand, provides physiological and environmental situation recordings collected via smart wristband devices.

These biological data can serve as the induced emotion labels, e.g. EEG signals can be translated into human emotions \cite{rahman2021recognition}. Increased datasets with different types of physiological signals enable researchers to focus on creating models for induced versus perceived emotion detection. This offers new avenues to use biofeedback in music emotion mediation applications through smart devices. 

Sometimes other emotion-inducing modalities are present, such as video, or lyrics. In this case, it is important to consider the influence of each of these modalities. In the case of video and music, Chua et al. \cite{chua2022predicting} have studied the influence of exposing participants not only to the music but also the muted video, as well as the music videos. They found that the music modality explains most of the variance in arousal values, and both music and video modalities explain the variance in valence values. Phuong et al. \cite{thao2021attendaffectnet} explore the influence of using only audio features and only video features to predict emotions from movie fragments. They found that the prediction is most accurate when both modalities are used. However, when predicting from a single modality, the audio model is most accurate. 

\textbf{Real-time} Many of the currently existing models are not implemented as an easy-to-use library, nor are they quick to run. They often require a GPU and may take several minutes to run. There are use cases, however, for real-time emotion recognition systems, as they would be able to integrate into therapeutic emotion detection systems \cite{agres2021music}, mood guidance playlist systems, as well as more commercial systems such as advertisement targeting systems. 

\textbf{Toward reliable and comparable MER research}
Despite significant advancements, the field still lacks standardized, high-quality datasets that serve as universal benchmarks. As discussed above, most existing datasets differ widely in genre coverage, emotion representation models, annotation procedures, and data quality. This heterogeneity, combined with inconsistent use of evaluation metrics and train/test splits, often makes it difficult to compare results across studies—even when using the same dataset. For example, the MTG-Jamendo dataset has been used in numerous studies (see Table~\ref{tab:models}), yet reported results vary substantially due to differing preprocessing strategies, loss functions, or model inputs. This wide variance raises questions about reproducibility and comparability in the field.

While recent initiatives like MusAV \cite{bogdanov2022musav} and unified multitask frameworks \cite{kang2025towards} represent important progress, the lack of standardized benchmarks still hinders progress. We argue that no single dataset can serve all research needs, especially given the subjective and culture-specific nature of emotion. Rather than searching for a one-size-fits-all dataset, we encourage the development of dataset-agnostic benchmarking tools, clearly defined splits, and model evaluation protocols. The community would benefit from centralized efforts (e.g., leaderboards, open splits, and documentation hubs) that promote transparency and make it easier to assess which models and results should be taken more seriously. Without such measures, the field risks being undermined by non-comparable results and hard-to-replicate studies.

In sum, the challenges mentioned above provide direct opportunities and future directions to further advance the exciting field of music emotion recognition. 

\section{Conclusion}
\label{sec:7}
Music emotion recognition is a promising field with various practical applications. With the rise of large-language models, we have seen impressive performance in various tasks. The field of music emotion recognition, however, still seems to be lagging. Given the importance of large training datasets to facilitate the training of LLMs, we provide a comprehensive overview and discussion of the existing datasets for music emotion recognition. 

We also explore current state-of-the-art models and dive into evaluation methods such as metrics as well as competitions, leaderboards, and benchmarks within the MER field. With this knowledge, we discuss the current challenges of the MER field at length and provide concrete future directions and emerging trends such as real-time systems and multimodal prediction systems. 

In closing, this survey serves as a valuable resource for the MER community, by offering insights into the current state-of-the-art, as well as a discussion of challenges and inspiration for future directions. 

\section{Acknowledgements}
This work has received SEED funding from SUTD TL under grant number RTDS S 22 14 04 1 and SUTD SKI 2021\_04\_06. 

\bibliographystyle{unsrt}
\bibliography{taffc}

\begin{IEEEbiography}[{\includegraphics[width=1in,height=1.25in,clip,keepaspectratio]{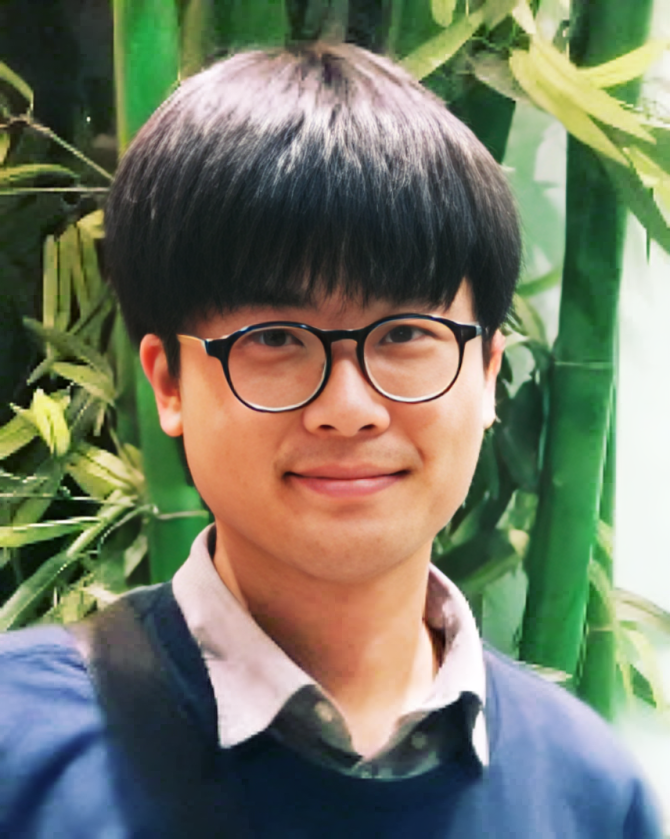}}]
{Jaeyong Kang} is a Postdoctoral Research Fellow at the Singapore University of Technology and Design (SUTD). He earned his Ph.D. in Electrical Engineering and Computer Science from the Gwangju Institute of Science and Technology (GIST), South Korea, in 2017. From 2018 to 2019, he served as a Research Scientist at the Biomedical Research Institute, Seoul National University Hospital (SNUH). His research interests span a diverse range of fields, including music generation, affective computing, deep learning, computer vision, natural language processing, agent-based information retrieval, social media analysis, and recommender systems.
\end{IEEEbiography}

\begin{IEEEbiography}[{\includegraphics[width=1in,height=1.25in,clip,keepaspectratio]{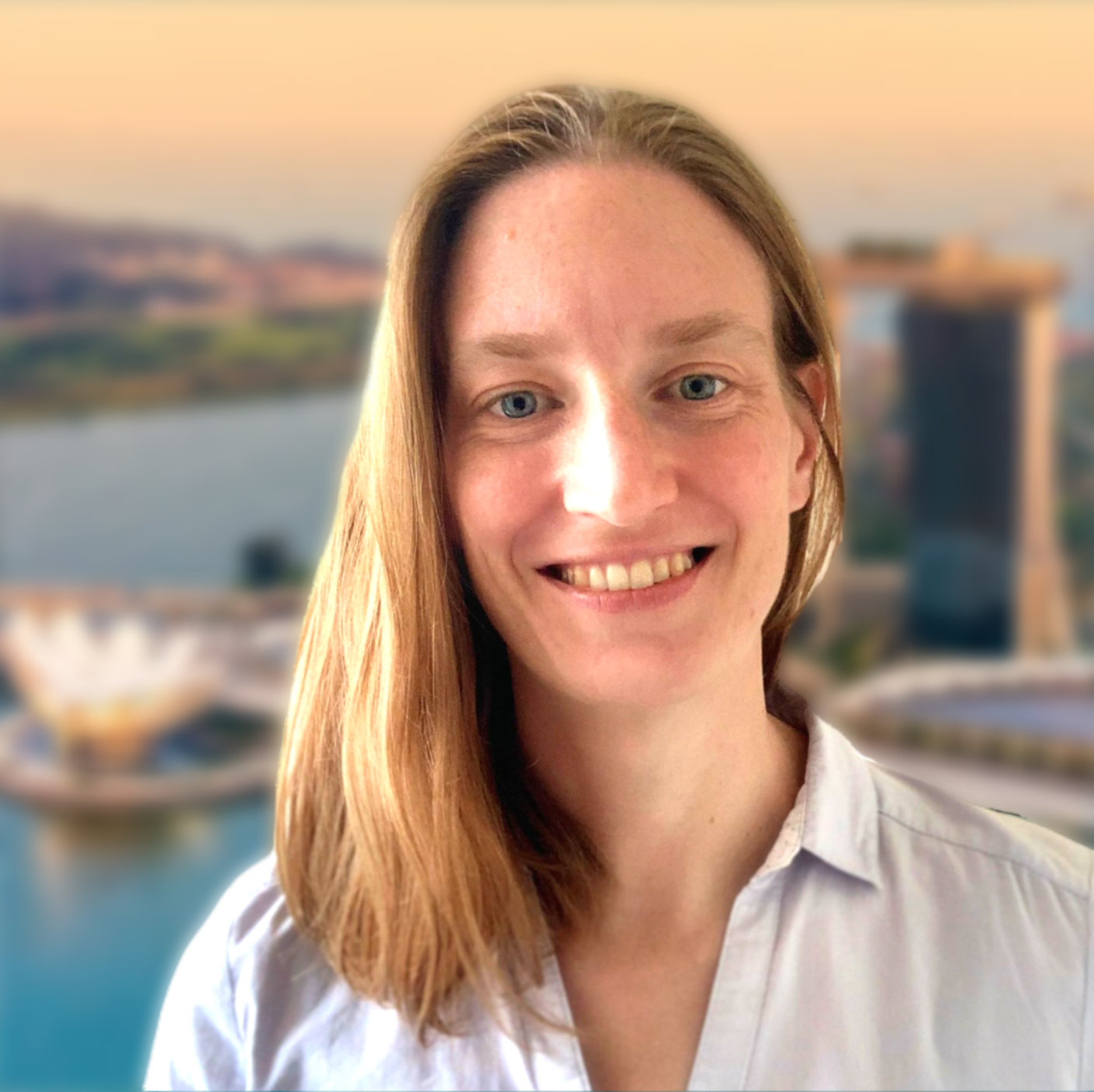}}]
{Dorien Herremans} is an Associate Professor at Singapore University of Technology and Design (SUTD), where she leads the Audio, Music, and AI (AMAAI) Lab. Her research focuses on developing cutting-edge AI technologies for multimodal applications, with a focus on generative models and affective computing. Before joining SUTD, she was a Marie Sklodowska-Curie Postdoctoral Fellow at the Centre for Digital Music at Queen Mary University of London.
Prof. Dorien Herremans has been a pioneer in the music technology field, publishing her first generative music model 20 years ago. 

She was also nominated on the Singapore 100 Women in Tech list in 2021, and shortlisted as one of the top 30 SAIL Award (Super AI Leader) Finalists in 2024 at the World AI Conference.

\end{IEEEbiography}
\vfill
\end{document}